\documentclass[a4paper,11pt]{article}
\usepackage{pos}
\usepackage{bibentry}

\title{Implications of Rare Kaon Decays on Lepton Number Violating Interactions}
\author[a]{Frank F. Deppisch}
\author*[b]{K\aa{}re Fridell}
\author[b]{Julia Harz}
\affiliation[a]{Department of Physics and Astronomy, University College London,\\
	London WC1E 6BT, United Kingdom}
\affiliation[b]{Physik Department T70, Technische Universität München,\\
	James-Franck-Straße 1, 85748 Garching, Germany}
\emailAdd{kare.fridell@tum.de}

\abstract{We explore the possibility of lepton number violation (LNV) manifesting in the rare kaon decay $K\to\pi\nu\nu$, and its consequences for radiative neutrino mass generation and the washout of Lepton asymmetry in high-scale leptogenesis scenarios. We perform the analysis in a model-independent framework, the Standard Model effective field theory (SMEFT), and discuss the possible LNV nature of the rare kaon decay in the context of the currently operating NA62 experiment at CERN. We find that, in case of a LNV interaction, its detection would put high-scale leptogenesis under tension and would hint to small radiatively generated neutrino masses.}

\FullConference{%
  40th International Conference on High Energy physics - ICHEP2020\\
  July 28 - August 6, 2020\\
  Prague, Czech Republic (virtual meeting)
}

\begin{document}
\maketitle

\section{Introduction}

Lepton number is the charge of an accidental symmetry of the Standard Model (SM) which is anomalously broken at the quantum mechanical level \cite{Klinkhamer:1984di}. This breaking, which occurs via sphaleron transitions, allows for an asymmetry in the lepton sector to be translated into a baryon asymmetry of the Universe (BAU). Such a mechanism could be one of the possibilities to generate the observed BAU \cite{Fukugita:1986hr}. 
Furthermore, the discovery of neutrino masses provides evidence of the need for New Physics (NP) beyond the SM (BSM), but the nature by which neutrinos acquire their mass remains unknown. If they, like the other fermions of the SM, receive their masses via a coupling to the Higgs boson, the value of the corresponding Yukawa coupling would be of the order $10^{-12}$. This would be significantly smaller than any Yukawa coupling in the SM, out of which the smallest is that of the electron at $y_e \approx 2.9\times 10^{-6}$ \cite{Tanabashi:2018oca}. Another way in which neutrino masses could be realized is through the extension of the SM Lagrangian by a Majorana mass term for the neutrino, and in this way, the symmetry which conserves lepton number is explicitly broken, possibly providing a link between the mass generation mechanism of the neutrino and the BAU. Majorana mass terms can be achieved in different ways \cite{Cai:2017jrq}, possibly through higher order processes involving additional leptons or quarks. In the latter case, it is possible that the corresponding NP degrees of freedom could be most stringently experimentally constrained in the LNV rare kaon decays $K\to\pi\nu\nu$ \cite{Deppisch:2020oyx}.

\section{Lepton number violation in rare kaon decays}

We investigate the effects of lepton number violating (LNV) interactions in the rare decays $K^+\to\pi^+ + \nu\bar\nu/\nu\nu$ and $K_L\to\pi^0 + \nu\bar\nu/\nu\nu$ in a model independent way using SM effective field theory (SMEFT). In this way, we extend the SM Lagrangian by effective operators with odd mass dimension $D\geq 5$ that violate lepton number by two units. Effects from high energy degrees of freedom are encoded in an EFT scale $\Lambda$, which, with a power $4-D$, acts as coefficients to the EFT operators in the Lagrangian. Operators with a low mass dimension will have an associated EFT scale with a smaller exponent as compared to operators with a high mass dimension. 
The lowest mass dimension at which the rare kaon decays can be mediated at short range is mass dimension 7, via the operator
\begin{equation}
\label{eq:op3b}
\mathcal{O}_{3b} = L^\alpha L^ \beta Q^\rho d^cH^\sigma\epsilon_{\alpha \rho}\epsilon_{\beta \sigma}.
\end{equation}
Here $L^\alpha$, $Q^\alpha$ and $H^\alpha$ are the $SU(2)_L$ doublet lepton, quark and Higgs fields, and $d^c$ is the down-type quark singlet. By short range, we mean that there are no vertices external to the operator that are needed in order to complete the diagram. This operator contributes to the Majorana mass $m_\nu$ of the neutrino via \cite{Deppisch:2020oyx}
\begin{equation}
\label{eq:numass_op3b}
\delta m_\nu^{(3b)} \approx \frac{y_d}{16\pi^2}\frac{v^2}{\Lambda_{3b}},
\end{equation}
where $y_d$ is the Yukawa coupling of a down-type quark, $v$ is the Higgs vacuum expectation value (vev), and $\Lambda_{3b}$ is the NP scale corresponding to operator $\mathcal{O}_{3b}$. Due to the observed neutrino mass being very small, Eq.~\eqref{eq:numass_op3b} puts tight constraints on the scale of operator $\mathcal{O}_{3b}$. For a mass $m_{\nu} = 0.1$~eV, the corresponding limit on the scale is $\Lambda_{3b} \geq 5.2\times 10^4$ TeV \cite{Deppisch:2020oyx}. However, this limit, obtained from an EFT operator, can have non-trivial corrections coming from a UV-completion, and should at most be taken as an estimate.

Due to the LNV nature of the operator $\mathcal{O}_{3b}$, it will contribute to a washout effect in a given leptogenesis scenario. Assuming an asymmetry to have been generated in the lepton sector at a high scale $T$, such as for example $T \gtrsim 10^{9}$ GeV as in a typical Type-I see-saw leptogenesis scenario \cite{Davidson:2002qv}, any LNV washout process that is active at a lower scale will act to reduce the generated asymmetry. If the scale of a LNV operator is found to be low relative to a possible scale of asymmetry generation, the washout coming from the operator will be strong over a long range in temperature, effectively removing any asymmetry, and therefore reducing the attractiveness of such a leptogenesis model. For this reason, if an observation of a LNV process is made at a low scale, it may lead to falsification of high scale leptogenesis scenarios \cite{Deppisch:2020oyx,Deppisch:2017ecm,Deppisch:2015yqa,Deppisch:2013jxa}. In order to connect the scale of LNV NP to a washout scale, we use a simplified Boltzmann equation \cite{Deppisch:2020oyx,Deppisch:2017ecm,Deppisch:2015yqa}
\begin{equation}
\label{eq:boltzmann}
\frac{d\eta_{\Delta L}}{dz} = - \frac{\eta_{\Delta L}}{z}c'_D\frac{\Lambda_{\text{Pl}}}{\Lambda_{3b}}\left(\frac{T}{\Lambda_{3b}}\right)^{5},
\end{equation}
where $\eta_{\Delta L}$ is the difference in number density between the $SU(2)_L$ lepton doublet $L^\alpha$ and the anti-lepton doublet $\bar L^\alpha$, normalized to the photon number density. Furthermore, $\Lambda_\text{Pl}$ is the Planck scale, $T$ is the temperature of the Universe, $z\propto T^{-1}$ is a time evolution variable and $c_D' = \frac{20\sqrt{195}}{56\pi^5\sqrt{107}}$. By solving Eq.~\eqref{eq:boltzmann} for a given NP scale $\Lambda_{3b}$, we obtain a temperature at which the washout stops being effective. Taking this scale as the lower limit of the washout, and the scale $\Lambda_{3b}$ as the upper limit, a temperature range is obtained for which the washout is effective.

\begin{figure}
	\centering
	\includegraphics[width=0.6\textwidth]{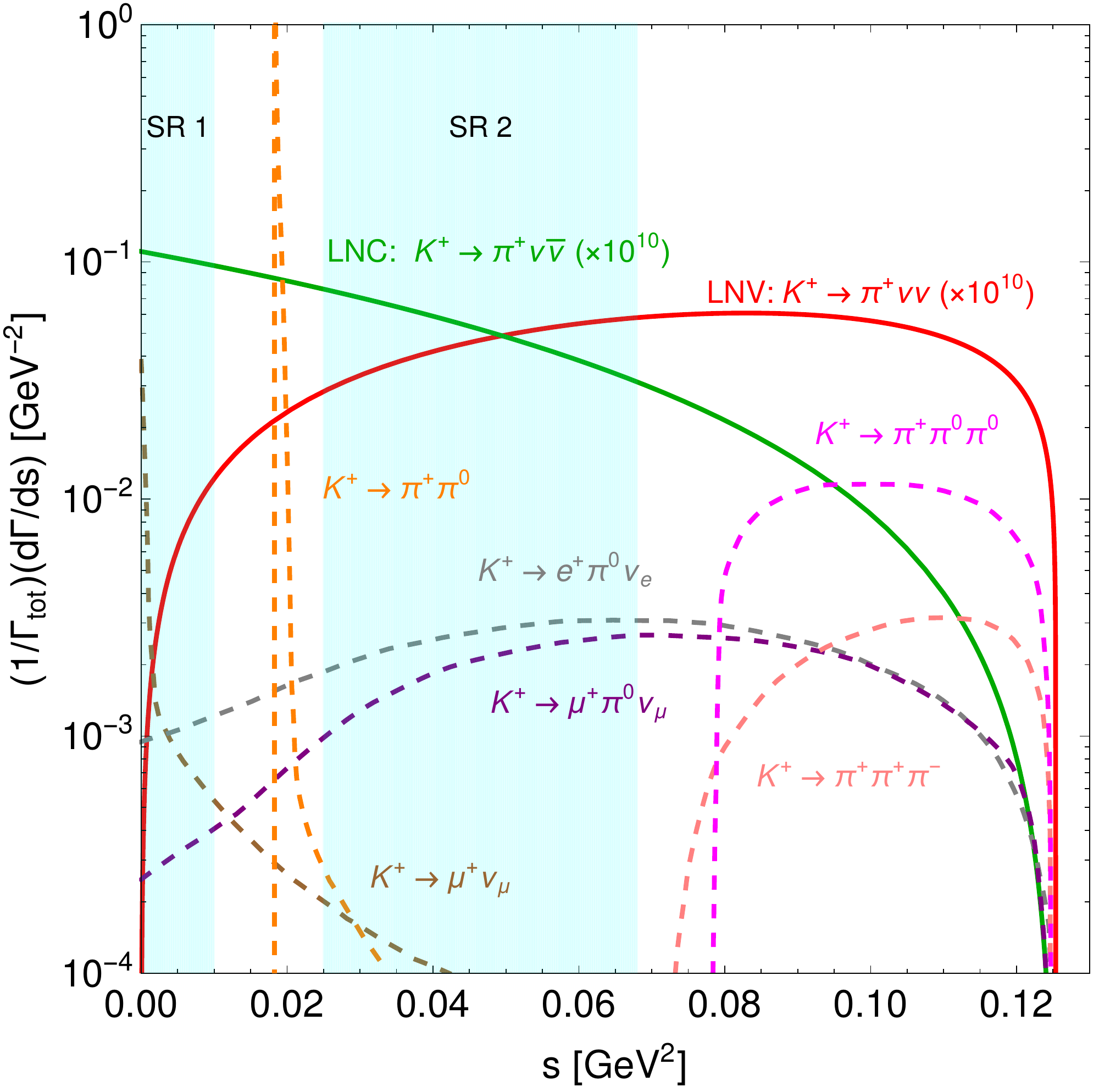}
	\caption{Kinematic distributions for different decay modes of charged kaons at the NA62 experiment \cite{CortinaGil:2020vlo}. Here the normalized differential decay width for a given mode is plotted against the squared missing energy~$s$. The two rare kaon decay modes $K^+\to\pi^+ + \nu\bar\nu/\nu\nu$ have been multiplied by a factor $10^{10}$ for visibility, all other decay modes are treated as background processes. The two shaded areas correspond to the two signal regions of the NA62 experiment.}
	\label{fig:distributions}
\end{figure}

From the operator $\mathcal{O}_{3b}$ in Eq.~\eqref{eq:op3b}, we obtain the squared matrix element for the LNV rare kaon decay $K(p_1)\to\pi(p_2)\nu(k_1)\nu(k_2)$, where $p_i$ and $k_i$ are the four-momenta of the particles, as
\begin{equation}
\label{eq:mat2BSM}
|\mathcal{M}(s)_\text{LNV}|^2 = \frac{v^2}{\Lambda^6_{3b}}
\left(\frac{m_K^2 - m_\pi^2}{m_s - m_d}f^K_0(s)\right)^2 s.
\end{equation}
Here $m_K$ and $m_\pi$ are the masses of $K^+/K_L$ and $\pi^+/\pi^0$ respectively for the charged$/$neutral decay mode, $m_s$ and $m_d$ are the masses of the strange and down quarks, $s=(p_1-p_2)^2$ is the square of the difference in energy between the kaon and the pion (which will also be referred to as "missing energy"
), and $f^K_0(s)$ is a form factor which is given by \cite{Mescia:2007kn}
\begin{equation}
\label{eq:formfactorzero}
f^K_0(s) = f^K_+(0)\left(1 + \lambda_0\frac{s}{m_\pi^2}\right),
\end{equation}
where $\lambda_0 = 13.38\times 10^{-3}$ and
\begin{equation}
\label{eq:fK0}
f^{K^+}_+(0) = 0.9778, \quad f^{K_L}_+(0) = 0.9544.
\end{equation}
In the lepton number conserving (LNC) decay, we get the following squared matrix element \cite{Deppisch:2020oyx}
\begin{equation}
\label{eq:mat2SM}
|\mathcal{M}(s,t)_\text{SM}|^2 = \frac{6}{\Lambda_\text{SM}^4}
\left[m_K^2\left(t-m_\pi^2\right) - t\left(s + t - m_\pi^2\right)\right]f^K_+(s)^2.
\end{equation}
Here $t = (p_2 + k_2)^2$, $\Lambda_\text{SM}$ is the effective scale of the process in the SM, and $f^K_+(s)$ is a form factor,
\begin{equation}
\label{eq:formfactorplus}
f^K_+(s) = f^K_+(0)\left(1 + \lambda_+'\frac{s}{m_\pi^2} + \lambda_+''\frac{s^2}{m_\pi^4}\right),
\end{equation}
with $\lambda_+' = 24.82\times10^{-3}$ and $\lambda_+'' = 1.64\times10^{-3}$. A comparison between the LNV and SM squared matrix elements from Eqs.~\eqref{eq:mat2BSM} and \eqref{eq:mat2SM} respectively can be made though the double differential decay width \cite{Deppisch:2020oyx}
\begin{equation}
\label{eq:decayrate}
\frac{\Gamma\left(K\to\pi\nu_i\nu_j\right)}{ds\,dt} =
\frac{1}{1+\delta_{ij}}\frac{1}{(2\pi)^3}\frac{1}{32m_K^3} |\mathcal{M}(s,t)|^2,
\end{equation}
where $i, j$ denote the neutrino flavours. From Eq.~\eqref{eq:decayrate}, a difference between the LNV and SM decay processes is apparent in the distribution of the decay rate. This difference is shown in Fig.~\ref{fig:distributions} where the differential decay widths of the LNV and SM rare kaon decays are shown, as well as several background processes in the NA62 experiment.

In Fig.~\ref{fig:distributions}, the normalized differential decay width\footnote{This distribution can be obtained by integrating Eq.~\eqref{eq:decayrate} over $t$.} $(1/\Gamma_\text{tot})d\Gamma/ds$ for different decay modes of $K^+$ at the NA62 experiment, with an incoming $K^+$ energy of 75 GeV, is shown as a function of the squared missing energy~$s$. Green and red solid lines show the distribution in $s$ corresponding to the LNC SM decay mode $K^+\to\pi^+\nu\bar\nu$ and the LNV BSM mode $K^+\to\pi^+\nu\nu$ respectively. Both of these rates have been multiplied by a factor $10^{10}$ for visibility. In shaded teal areas, the two signal regions SR1 and SR2 are displayed, which correspond to the range in $s$ for which candidate events for the NA62 experiment are accepted. These regions are placed such that the greatest part of the two-pion and three-pion final state modes are avoided. As evident in Fig.~\ref{fig:distributions}, the distribution of the SM LNC decay mode peaks in SR1, and subsequently falls off for higher values of $s$. In contrast, for the distribution corresponding to the LNV mode, SR1 poses a very small event detection prospect, and instead, higher values of $s$ are more favored, with a peak being present around $s\approx 0.08$ GeV. As such, SR2 would be expected to receive a comparable number of events for both modes, while a signal in SR1 plainly advocates the SM LNC decay.

\begin{table}
\centering
\begin{tabular}{l l l}
	\hline\hline
	Signal region & Vector current & Scalar current\\
	\hline
	NA62 SR 1 & 6\% & 0.3\%\\
	NA62 SR 2 & 17\% & 15\%\\
	\hline\hline
\end{tabular}
\caption{Percentage of the phase space distribution of rare kaon decays proceeding via vector or scalar currents that falls within the corresponding signal regions of the NA62 experiment.}
\label{tab:percentages}
\end{table}

Tab.~\ref{tab:percentages} shows the percentages of total phase space falling within the different signal regions of the NA62 experiment for the SM LNC and BSM LNV decay modes both. As can also be inferred from Fig.\ref{fig:distributions}, Tab.~\ref{tab:percentages} shows that events in SR1 are more favorably attributed to the SM decay, and that a similar preference is given for both modes in SR2. With a large set of data, it is conceivable that a determination of the LNV/LNC nature of the rare kaon decay may be produced, by accounting for the difference in number of events in the two signal regions. However, considering the small branching ratio of the rare kaon decay, such a large data set would be very difficult to construct.

\section{Results}
By evaluating the decay width for the LNV rare kaon decay $K^+\to\pi^+\nu\nu$ while taking into account the difference in signal acceptance to that of the LNC SM decay, the corresponding lower limit on the EFT scale of operator $\mathcal{O}_{3b}$ is calculated. With the current and future experimental limits on the branching ratio from NA62 being $\text{BR}^\text{NA62}_\text{current}<1.78\times 10^{-10}$ \cite{CortinaGil:2020vlo} and $\text{BR}^\text{NA62}_\text{future}<1.11\times 10^{-10}$ respectively, at 90\% confidence level, the NP scale limits are obtained as $\Lambda^\text{NP}_\text{current}>17.2$~TeV and $\Lambda^\text{NP}_\text{future}>19.6$~TeV \cite{Deppisch:2020oyx}, respectively. The effect on leptogenesis, as evaluated using Eq.\eqref{eq:boltzmann}, of a LNV NP scale residing close to these limits, is a washout that is highly effective down to temperatures of 196~GeV and 213~GeV respectively for the current and future experimental limits~\cite{Deppisch:2020oyx}. In both cases, the washout is effective almost all the way down to the electroweak symmetry breaking scale 174~GeV, at which point the sphaleron interaction shuts of, and baryogenesis via leptogenesis is no longer possible in the conventional form.

\section{Summary}

Using an EFT framework, we connect the current and future experimental limits on rare kaon decays to radiative neutrino mass generation, as well as to the washout of lepton asymmetry in high scale leptogenesis scenarios. 
We show that if LNV is realized close to current experimental bounds in rare kaon decays, high-scale leptogenesis would be put under significant pressure, as the washout of lepton asymmetry would remain strong close to the EW symmetry breaking scale. Furthermore, we suggest a plausible way to distinguish the LNV decay $K^+\rightarrow\pi^+\nu\nu$ from its LNC SM counterpart $K^+\rightarrow\pi^+\nu\bar{\nu}$ at the NA62 experiment, by acknowledging the difference in phase space distribution of the final pion between decays mediated by scalar and vector currents, which corresponds to a difference in chirality of the two final neutrinos. Searching for LNV in nature is of dire importance in order to answer several open questions in particle phenomenology, such as the origin of neutrino masses and the baryon asymmetry of the Universe. Mainly because of its theoretical cleanliness, the rare kaon decay $K\rightarrow\pi\nu\nu$ provides an excellent probe of BSM physics, and as such, future experimental results could decide whether NP is realized at an observable energy scale.

\acknowledgments
We thank Chandan Hati for helpful discussions. F.F.D. was supported by the UK Science and Technology Facilities Council (STFC) via a Consolidated Grant (Reference ST/P00072X/1). K.F. and J.H. acknowledge support from the DFG Emmy Noether Grant No. HA 8555/1-1.


\begin{thebibliography}{99}

\bibitem{Klinkhamer:1984di}
F.~R.~Klinkhamer and N.~S.~Manton,
Phys. Rev. D \textbf{30} (1984), 2212
doi:10.1103/PhysRevD.30.2212

\bibitem{Fukugita:1986hr}
M.~Fukugita and T.~Yanagida,
Phys. Lett. B \textbf{174} (1986), 45-47
doi:10.1016/0370-2693(86)91126-3

\bibitem{Tanabashi:2018oca}
M.~Tanabashi \textit{et al.} [Particle Data Group],
Phys. Rev. D \textbf{98} (2018) no.3, 030001
doi:10.1103/PhysRevD.98.030001

\bibitem{Cai:2017jrq}
Y.~Cai, J.~Herrero-Garc\'\i{}a, M.~A.~Schmidt, A.~Vicente and R.~R.~Volkas,
Front. in Phys. \textbf{5} (2017), 63
doi:10.3389/fphy.2017.00063
[arXiv:1706.08524 [hep-ph]].

\bibitem{Deppisch:2020oyx}
F.~F.~Deppisch, K.~Fridell and J.~Harz,
[arXiv:2009.04494 [hep-ph]].

\bibitem{Davidson:2002qv}
S.~Davidson and A.~Ibarra,
Phys. Lett. B \textbf{535} (2002), 25-32
doi:10.1016/S0370-2693(02)01735-5
[arXiv:hep-ph/0202239 [hep-ph]].

\bibitem{Deppisch:2017ecm}
F.~F.~Deppisch, L.~Graf, J.~Harz and W.~C.~Huang,
Phys. Rev. D \textbf{98} (2018) no.5, 055029
doi:10.1103/PhysRevD.98.055029
[arXiv:1711.10432 [hep-ph]].

\bibitem{Deppisch:2015yqa}
F.~F.~Deppisch, J.~Harz, M.~Hirsch, W.~C.~Huang and H.~P\"as,
Phys. Rev. D \textbf{92} (2015) no.3, 036005
doi:10.1103/PhysRevD.92.036005
[arXiv:1503.04825 [hep-ph]].

\bibitem{Deppisch:2013jxa}
F.~F.~Deppisch, J.~Harz and M.~Hirsch,
Phys. Rev. Lett. \textbf{112} (2014), 221601
doi:10.1103/PhysRevLett.112.221601
[arXiv:1312.4447 [hep-ph]].

\bibitem{Mescia:2007kn}
F.~Mescia and C.~Smith,
Phys. Rev. D \textbf{76} (2007), 034017
doi:10.1103/PhysRevD.76.034017
[arXiv:0705.2025 [hep-ph]].

\bibitem{CortinaGil:2020vlo}
E.~Cortina Gil \textit{et al.} [NA62],
JHEP \textbf{11} (2020), 042
doi:10.1007/JHEP11(2020)042
[arXiv:2007.08218 [hep-ex]].
\end{thebibliography}
\end{document}